# Barrier Breakdown in a Multiple Quantum Well Structure

A. Gomez[1, 2] and V. Berger[1, 2]

[1]Laboratoire MPQ, Université Paris 7, CNRS, Case 7021, 75205 PARIS Cedex 13, FRANCE and

[2]Alcatel-Thales 3-5 lab, Route départementale 128, 91767 PALAISEAU Cedex, FRANCE

N. Péré-Laperne[3] and L.A. De Vaulchier[3]

[3]Laboratoire Pierre Aigrain, CNRS, Ecole normale supérieure, 24 Rue Lhomond, 75005 PARIS Cedex, FRANCE

We explore a regime of unipolar electronic transport in a multiple quantum well structure with very large current discontinuities - up to five orders of magnitude. Magneto-transport experiments reveal different transport regimes. Quantum well impact ionization shifts the structure from a resistive "down" state, where the current flows through inter-well quantum tunneling, to a highly conductive "up" state. In the latter regime, the current leaks through a barrier suddenly broken down because of an efficient ionization of the first quantum well.

72.20.-i

73.63.Hs

85.30.-z

Large current discontinuities and instabilities have been observed in different types of semiconductor heterostructures. This has illustrated fundamental aspects such as quantum tunneling and has given birth to many important devices from detectors to microwave sources[1]. Different physical mechanisms are usually responsible for current discontinuities: thermal instabilities[2], tunneling[3,4], or band to band impact ionization[5]. Impact ionization has also been observed in unipolar n-GaAs bulk materials[6], with nonlinear and hysteretic current voltage characteristics. In this letter, we focus on multiple quantum well structures and impact ionization is found to be responsible for huge current discontinuities up to several orders of magnitude. The different regimes of electronic transport are analyzed through magneto-transport experiments. We anticipate this observation to be of great use for the development of original devices such as very high frequency microwave generators or Thz avalanche detectors.

Sudden current increases are observed in a multiple quantum well structure. During this transition from a resistive to a conductive state, the current is increased by more than five orders of magnitude. This occurs when the voltage reaches a threshold where the first quantum well is depleted by efficient impact ionization. This depletion induces an enhancement of the local electric field and a barrier breakdown, and dramatically increases the current injection at the contact. Hysteresis is observed between these two states, empty well - large current (further referred to as "up" state) and full well - low current (further referred to as "down" state). Magneto-transport experiments emphasize the importance of the capture by impact ionization in this process. Electronic capture can indeed be controlled by the quantization into Landau levels, leading to a strong modulation of the transport properties in the structure.

The sample contains 30 GaAs quantum wells of thickness $L_w$ = 15.5 nm separated by 70.2 nm ($L_b$) wide $Al_{0.03}Ga_{0.97}As$ barriers. Each quantum well is n-doped in its center, 10 nm long, with a nominal doping value $n_{2D}$ = $6.10^{10}$ $cm^{-2}$. This structure is embedded between two contacts doped up to $10^{17} cm^{-3}$. The sample was initially purposed for infrared quantum well detection and grown by molecular beam epitaxy in the group of H.C. Liu at the National Research Council Canada. The performances of this detector (referred to as V266 in Ref.[7]) were fully characterized in Ref.[7]. The shallow quantum wells contain only one confined bound state (with an energy confinement $E_1$ of 8.6 meV). This leads to an infrared photodetector with a photoresponse peaked at a wavelength of 56 microns at an energy slightly above the difference between the fundamental level and the barrier, equal to 18 meV.

The I(V) characteristics at zero magnetic field and at a temperature of 1.5 K is shown in figure 1. At low bias, the current is classically attributed to direct tunneling between adjacent quantum wells. At such low temperature and bias, electrons do not have enough energy to escape from the quantum wells. Indeed, the Fermi level $E_{Fw}$ is equal to 2 meV above the ground state $E_1$, the thermal energy to 125 μeV and the energy decreases by only 600 eV/cm at 0.15V applied bias, which gives an energy drop of only 4 meV along one barrier width. These energies are small compared to the one of the continuum which is 18 meV above the fundamental level. When the bias is increased, at a threshold of 0.255 V, the current shows an abrupt increase of more than five orders of magnitude, reaching a typical value of 2.4 A/cm² at a voltage of 0.3 V. This remarkable transition of the system to an « up » state shows that the physical mechanism governing the electronic transport is totally different. In this new regime, the electronic injection in the structure is very efficient and implies a barrier breakdown at the contact. This barrier breakdown is attributed to impact ionization of the first QW, which is emptied of its electrons, resulting in an electric field discontinuity at the first QW and therefore a strong electric field upstream this well.

Let us first study the "down" state regime of inter-well tunneling. The current density can be described with a classical theory of electron elastic emission through a confinement barrier. It is given by the product $J_{down} = n_{2D} q \tau_{tunnel}^{-1}$, where the time $\tau_{tunnel}$ is classically related to the energy of electrons $E_1$, the quantum well thickness $L_w$, and the tunnel probability $P$ by $\tau_{tunnel} = (2L_w / v_1) P^{-1}$ (with $E_1 = (mv_1^2 / 2)$). The tunnel probability $P$ through the barrier separating the two quantum wells is given by:[8]

$$P = exp\left(-\frac{4}{3qF}\right)\sqrt{\frac{2m^*}{\hbar^2}} \times \left[(V_b - E_1)^{3/2} - (V_b - E_1 - qFL_b)^{3/2}\right] \quad (1)$$

The above expression of $J_{down}$ gives the contribution of the current flowing from a QW to a lower energy adjacent QW, where an electric field F is applied between them. $V_b$ is the barrier height.

The total current expression also needs to take into account the current coming backwards from the lower energy quantum well. This backward current is significant only for low electric fields (lower than 250V/cm) such that the electrons on the fundamental subband of the lower QW have enough energy to scatter elastically to the confined state of the upstream QW. The current density is finally given by:

$$J_{down} = \frac{qm^*k_BT}{\pi\hbar^2}\tau_{tunnel}^{-1} \log\left[\frac{1+\exp\left(\frac{E_{Fw}-E_1}{k_BT}\right)}{1+\exp\left(\frac{E_{Fw}-qFL_B-E_1}{k_BT}\right)}\right] \quad (2)$$

At such low temperature and low field (F < 250V/cm which corresponds to V < 65 mV), $k_BT \ll E_{Fw} - E_1$ yields a simplified expression of the down state current $J_{down} = (qm^*/\pi\hbar^2)\tau_{tunnel}^{-1}FL_b$ instead of $J_{down} = n_{2D}q\tau_{tunnel}^{-1}$ suitable at higher bias.

Above 250 V/cm, expression (2) has been used to approximate the current density in figure (1), with an excellent agreement. This validates the physical interpretation of this "down" regime.

Let us switch now to the description of the "up" state. The very high current density directly gives the value of the electric field $F_0$ on the injection barrier. This electric field is high enough for the electrons to tunnel through a triangular barrier (i.e. $F_0.L_b > 18$ meV). In this case, the relation between $F_0$ and the current density $J_{up}$ can be described by a WKB expression :

$$J_{up} = \frac{qm^*k_BT}{2\pi^2\hbar^3} \int_0^{V_b} exp\left(\left(-\frac{4}{3qF_0}\right)\sqrt{\frac{2m^*}{\hbar^2}}\left(V_b - E\right)^{3/2}\right) \times log\left[1 + exp\left(\frac{E_{Fc} - E}{k_BT}\right)\right] dE \quad (3)$$

with $E_{Fc}$, the Fermi level in the contact.

In this expression, the backward current has been neglected (consistently with the high value of the electric field). For example, with $j_{up}$=2.4 A/cm$^2$ at 0.3V, this expression leads to an electric field $F_0$ = 7.3 kV/cm, and to an energy drop of 51 meV along the first barrier.

This barrier breakdown can be explained by the following arguments: for a sufficiently high electric field, part of the electrons can escape and flow above the barrier. This induces impact ionization of the QW and further increases the number of electrons in the continuum. The effect of this impact ionization process is to deplete the QW, resulting in a positive charge in this QW, due to silicon donors. As a result, at this first QW, an electric field discontinuity $\Delta F$ proportional to the QW depletion, $\rho_S$, occurs according to Poisson relation $\Delta F = -e\rho_S /\varepsilon_0 \varepsilon_r$. This increase in the electric field further amplifies the current injection in the continuum according to Eq. (3), and induces more impact ionization. These impact ionization processes have already been observed in the context of quantum well infrared photodetectors (QWIPs)[9]. They are responsible for excess current and noise in these devices, especially in long wavelength detectors. In our case, the shallow nature of the quantum well and the very low temperature both dramatically enhance this effect, leading to the barrier breakdown.

This barrier breakdown occurs when the electric field is high enough to induce the first impact ionizations. The depletion of this QW is then abruptly completed by the catastrophic process: impact ionizations $\Rightarrow$ positively charged QW $\Rightarrow$ higher upstream electric field $\Rightarrow$ more electrons in the continuum $\Rightarrow$ impact ionizations. Through this transition, the QW has switched from a "full" state (represented in figure 2a for V = 0.15 V) to a quasi "empty" one (represented in figure 2b for V = 0.3 V).

This behavior is confirmed by the hysteresis in the I(V) curve, where the threshold going back from the "up" state to the "down" state is observed at a lower bias (0.164V). Once the QW is empty, the system is stable in the "up" state with the high electric field, and a lower bias is necessary to force the electric field to come back to a low value. The Poisson equation applied on the first QW driving the electric field distribution, together with the current-electric field relation (3), and the capture by impact ionization, have actually two stable "up" and "down" solutions for biases between 0.164 and 0.255 V: the state of the system depends on its history. The asymmetry of the I(V) between negative and positive bias is classical in the context of QWIPs and is due to silicon and aluminum segregation in the QW. This will be studied in further details, later.

These interpretations are confirmed by magnetic field experiments. Indeed, the transport of the confined electrons in a QW is modified by the quantization into Landau levels because of a magnetic field which modifies the density of states. As a consequence, the capture by an impact ionization process is strongly affected by this density of states. In the "down" state of the system, the magnetic field has also a strong impact on the current as will be discussed in the next paragraph.

The current as a function of the magnetic field is shown in figure 3 at 0.15V for a temperature of 1.5 K. Strong resonances are observed, and appear at magnetic fields such that different Landau levels are resonant with the barrier. These resonances are labeled on figure 3 with different Landau indices. For magnetic fields equal to 2.9, 4, 6.4 and 13.5 Tesla, the respective $|1,4>$, $|1,3>$, $|1,2>$ and $|1,1>$ Landau levels are resonant with the barrier (using the usual expression $E_{n,p} = E_n^0 + (1/2 + p)\hbar eB/m^*$).

When such a resonance occurs, some electrons can transfer to the high Landau level density of states, and escape to the continuum, leading to an additional current. This process to a quasi bound Landau level can be mediated through inelastic electron-electron scattering, which dominates at low temperature[10]. Interface roughness scattering may also take part to these electron transfers[11]. The population of electrons on these excited Landau levels at such a low temperature implies that the system remains in a non-equilibrium state characterized by an electronic temperature larger than the lattice temperature[12]. The density of states in Landau levels increases with the magnetic field, and results in a stronger current enhancement for $|1,1>$ resonances than for lower magnetic field resonances, in agreement with observations of figure 3. On the other side, when the magnetic field is such that no Landau level is resonant with the barrier, the current is lower because the final states for the electrons are either below the barrier (with a low escape probability) or above the barrier, and the energy necessary for inelastic scattering is higher, decreasing the probability of this process.

The current as a function of the magnetic field is shown in figure 4 with the multiple QW in the "up" state (V=0.3V, T=1.5K). The main effect of the magnetic field is to suppress the current, turning the sample back to the down state, with an hysteresis behavior. Once in the down state, the current as a function of the magnetic field shows the same resonances as described before (amplified by a factor 70000 in figure 4 for clarity). The position of the resonances between the Landau levels and the continuum are slightly shifted to lower magnetic field with respect to fig. 3 consistently with barrier lowering with the electric field. This "up" to "down" state transition of the multiple QW structure with the magnetic field confirms the role of impact ionization in the "up" state. Indeed, impact ionization strongly depends on the magnetic field: The density of states switches continuously from a continuum to a discrete structure. When this discrete structure is established (when the Landau level separation is greater than the finesse), the conservation of energy reduces drastically the number of electronic states suitable for an impact ionization. This relies on the same mechanism as the LO phonon relaxation bottleneck in quantum boxes[13] or in quantum wells under magnetic field[14]. Impact ionization disappears and finally switches the system back to the "down" state.

These transport regimes appear simultaneously in figure 5 which depicts the current as a function of the inverse of the temperature for two different values of the bias corresponding to the "down" (0.15 V) and "up" state (0.3 V). In the "up" state, the current remains stable with the temperature. As expected, no activation energy is associated to the huge tunneling current resulting from the injection barrier breakdown. The same behavior is found at low temperature (from 4.5 K to 10 K) for the "down" state: elastic inter-well tunneling does not depend on the temperature. At higher temperature (>10 K), dark transport evolves from an inter-well tunneling regime to a thermo-ionic one with a corresponding activation energy $E_a$ of 18 meV related to acoustic phonon absorption from the confined level $E_1$ to the continuum, as expected.

In conclusion, we have explored a new regime of electronic transport in unipolar devices. Spectacular barrier breakdown are attributed to impact ionization in a quantum well. Beyond the interest of this physical mechanism, we expect that this original effect can be used to design new devices such as very high frequency microwave generators, or THz avalanche unipolar detectors with the perspectives of photon counting in the ThZ range.

The authors are deeply indebted to H. C. Liu for providing the sample V266 of ref. 7, and also to M. Carras and Y. Guldner for fruitful discussions. One of the authors (A. G.) is supported by a DGA fellowship.

Correspondence and requests for materials should be addressed to vincent.berger@univ-paris-diderot.fr.

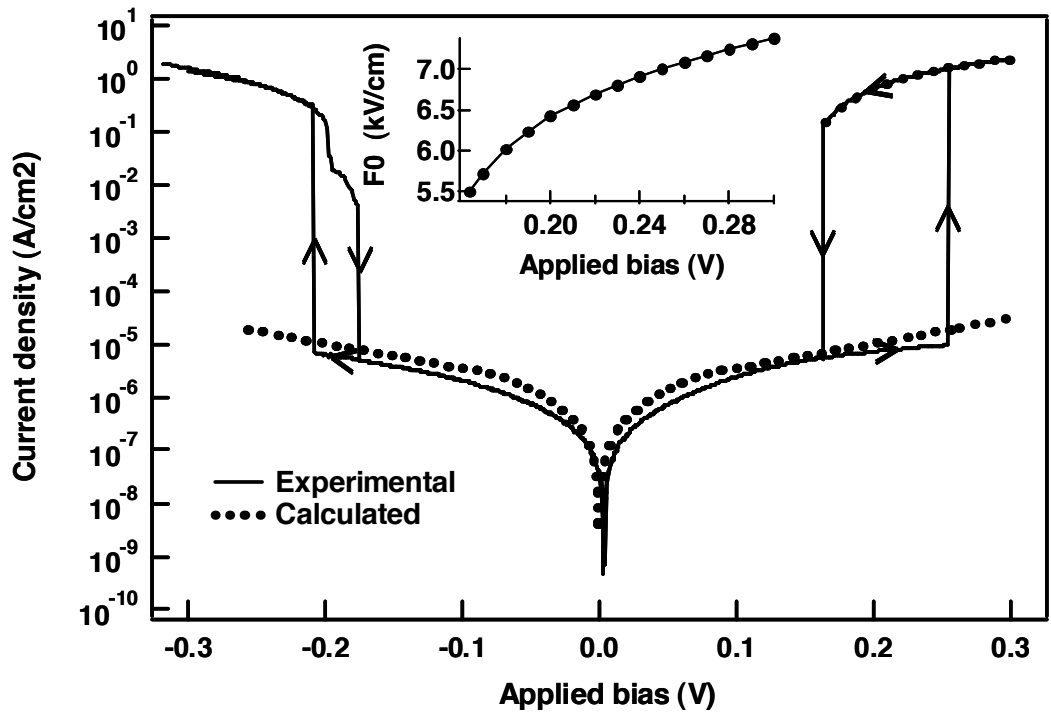

**Figure 1 : Current density as a function of the applied bias. The arrows indicate the increasing and decreasing bias paths. The dotted line is the modeling of the current in the "down" state given by Eq. (2) and in the up state given by Eq. (3). The inset shows the electric field $F_0$ extracted from the current Jup.**

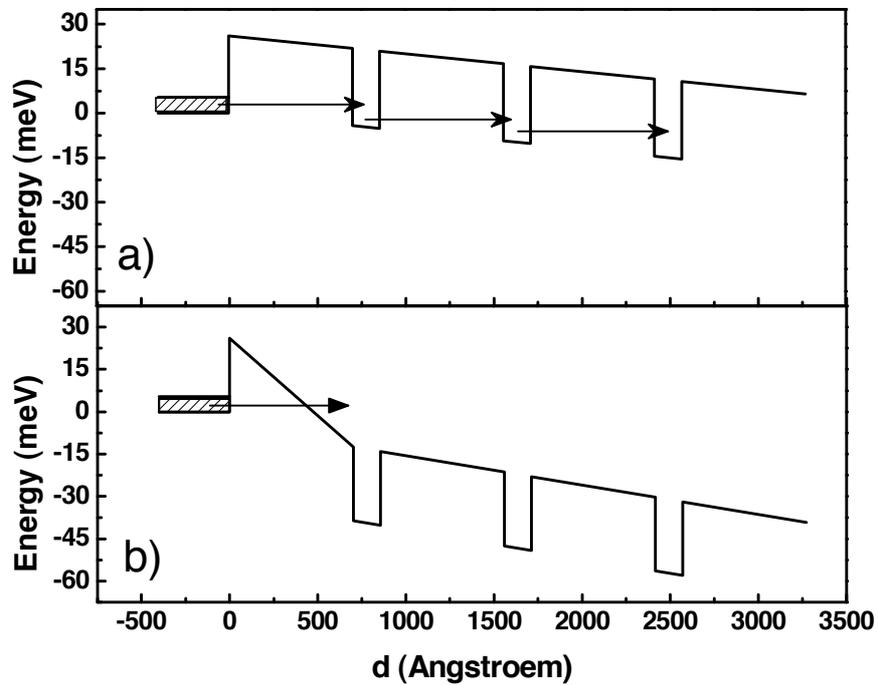

**Figure 2: Band structure of the multiple quantum well under applied bias. a) At 0.15V, in the "down" state, current is due to inter-well tunneling. b) At 0.3V, in the "up" state, efficient impact ionization is responsible for complete quantum well depletion, resulting in a high electric field applied on the injection barrier.**

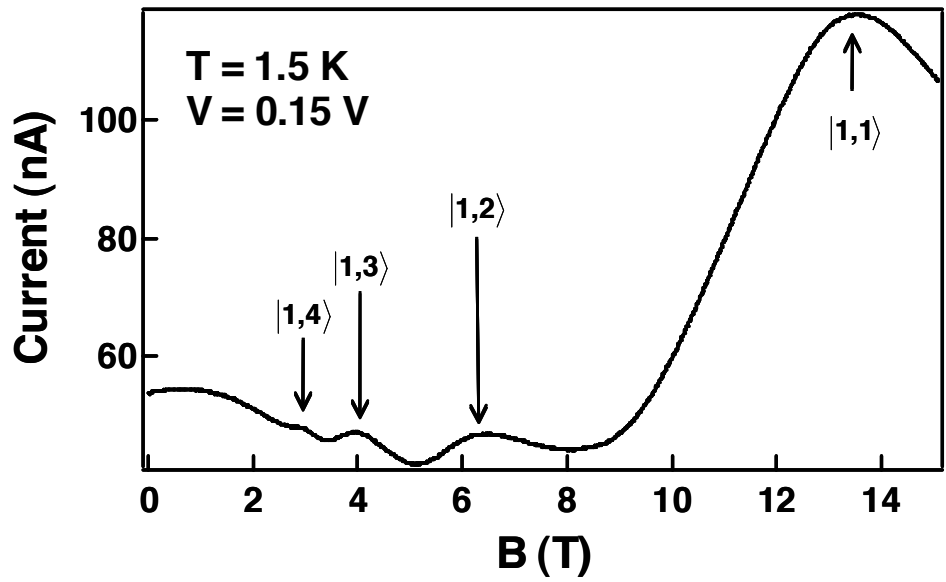

**Figure 3: Current in the "down" state as a function of the magnetic field. Resonances occur when a Landau level is resonant with the barrier**

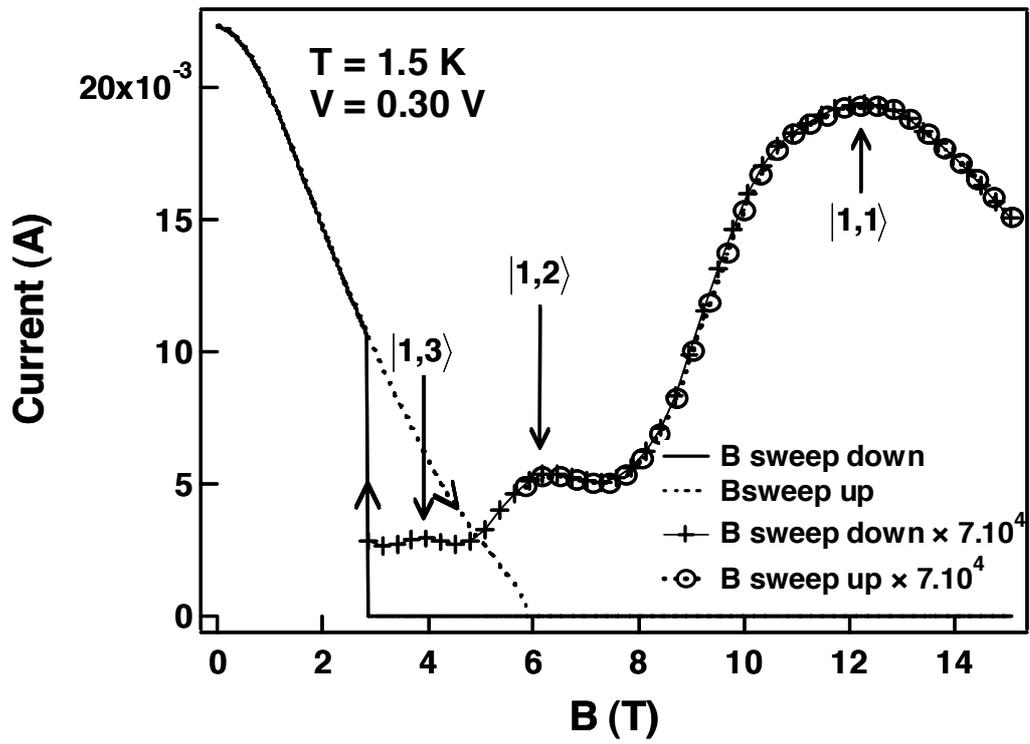

**Figure 4: Current switch off and on as a function of the magnetic field**

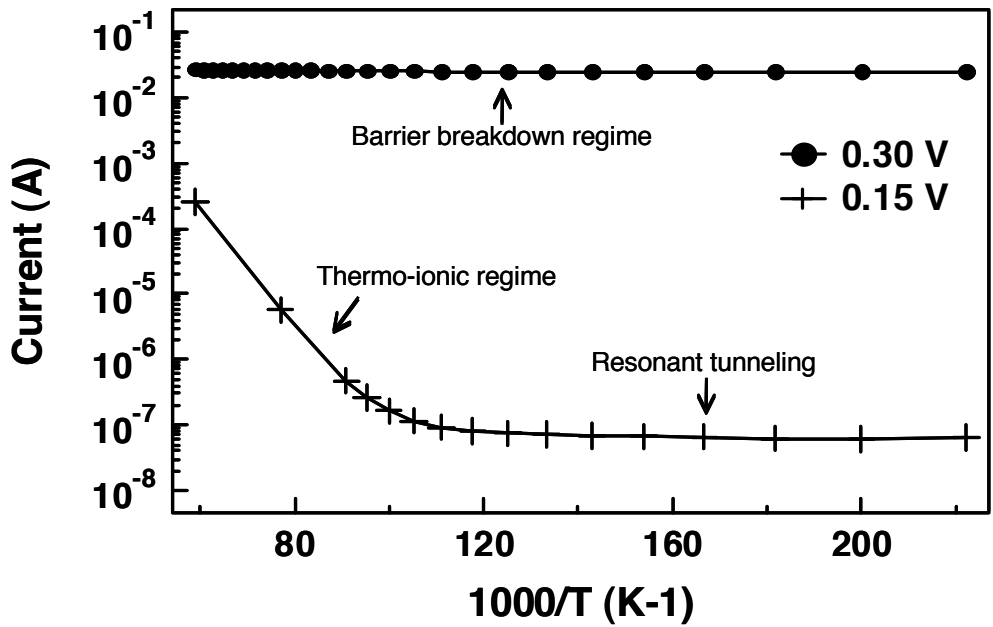

**Figure 5: Current as a function of the inverse temperature in the "down" and in the "up" states (crosses and circles, respectively).**